\documentclass[sigconf, nonacm]{acmart}

\usepackage[frozencache=true,cachedir=_minted-blobshuffle]{minted}
\usepackage[capitalise]{cleveref}
\usepackage{subcaption}
\usepackage[most]{tcolorbox}
\usepackage{tabularx}
\usepackage{enumitem}

\newtcolorbox{insetbox}[1][]{enhanced,boxsep=2pt,left=5pt,right=5pt,boxrule=\heavyrulewidth,arc=0pt,colframe=black, 
	colback=white,
	coltitle=black,
	,#1}

\newtcolorbox{resultsbox}[2][]{enhanced,boxsep=0pt,top=\belowrulesep,bottom=\aboverulesep,left=0pt,right=0pt,boxrule=0pt,arc=0pt,colframe=white, 
	colback=white,
	coltitle=black,
	title={#2},
	detach title,
	before upper={\strut\textbf{\tcbtitle}\quad},
	after upper={\strut},
	borderline north={-\heavyrulewidth}{0pt}{black, line width=\heavyrulewidth},
	borderline south={-\heavyrulewidth}{0pt}{black,line width=\heavyrulewidth}
	,#1}

\begin{document}
\title{BlobShuffle: Cost-Effective Repartitioning in Stream Processing Systems via Object Storage Exemplified with Kafka Streams}

\author{Sören Henning}
\orcid{0000-0001-6912-2549}
\affiliation{%
	\institution{Dynatrace Research}%
	\city{Linz}%
	\country{Austria}%
}
\email{soeren.henning@dynatrace.com}

\author{Otmar Ertl}
\orcid{0000-0001-7322-6332}
\affiliation{%
	\institution{Dynatrace Research}%
	\city{Linz}%
	\country{Austria}%
}
\email{otmar.ertl@dynatrace.com}

\author{Adriano Vogel}
\orcid{0000-0003-3299-2641}
\affiliation{%
	\institution{Dynatrace Research}%
	\city{Linz}%
	\country{Austria}%
}
\email{adriano.vogel@dynatrace.com}

\begin{abstract}
Shuffling or repartitioning data streams is an essential operation of state-of-the-art stream processing frameworks to support stateful workloads in a large-scale, distributed setting.
In today's cloud deployments, however, shuffling can become a major cost driver due to substantial network traffic across multiple availability zones (AZs) as well as an operational burden when operating a high-throughput, strongly consistent messaging backbone at scale.

We present BlobShuffle, a novel approach to cost-effective shuffling for stream processing systems that leverages cloud object storage as an intermediate exchange layer. Instead of sending all shuffled records directly, BlobShuffle groups records into batches, stores these batches in cloud object storage, and forwards only compact notifications. Downstream operators use these notifications to retrieve the relevant batches and extract the corresponding records. BlobShuffle balances cost efficiency and latency through configurable batching and a distributed caching mechanism.

BlobShuffle is implemented as an add-on for Kafka Streams that requires only minimal code changes to existing applications, leaves Kafka and the underlying infrastructure unmodified, and preserves Kafka Streams' consistency and correctness guarantees. In a large-scale experimental evaluation on a Kubernetes-based AWS deployment, we show that BlobShuffle can reduce shuffling costs by more than $40\times$ compared to native Kafka Streams shuffling while keeping the 95th percentile shuffle latency below 2\,seconds. Moreover, it scales to processing more than 2\,GiB/s without encountering a scalability limit in our experiments, indicating that BlobShuffle can economically support shuffle-intensive workloads at large scale.
\end{abstract}

\maketitle

\section{Introduction}

Stream processing has become a foundational technology for modern data-intensive systems, enabling near-real-time analysis of continuous data streams~\cite{Fragkoulis2023}. A key operation in many stream processing workloads is \emph{shuffling} (also called \emph{repartitioning}), where records are redistributed across parallel operators based on a key. Shuffling is required whenever operators group, aggregate, or join records by key and, thus, is essential for both correctness and scalability for stateful stream processing applications~\cite{ICPE2024}.

In cloud-based deployments, however, shuffling can become a major cost driver. To ensure fault tolerance, stream processing systems are typically deployed across multiple availability zones (AZs). As a result, repartitioning is responsible for continuous cross-AZ traffic, incurring significant charges on many cloud providers that often dominate the operational cost of stateful workloads~\cite{Merli2025}.

Kafka Streams~\cite{Sax2018,Wang2021} is a widely used, library-style stream processing framework that builds on Apache Kafka as its messaging and storage backbone. Similar to other frameworks, Kafka Streams relies on repartitioning to route records with the same key to the same operator instance. In Kafka Streams, shuffling is implemented via dedicated \emph{repartitioning topics} hosted by the Kafka brokers. Each shuffle step writes records to a repartitioning topic, from which downstream tasks then read.
While this design leverages Kafka's durability and fault tolerance, it amplifies two key challenges:
\begin{itemize}
	\item \textbf{Cloud costs.} Shuffling causes cross-AZ traffic when records are written to repartitioning topic partitions hosted in different AZs. In addition, these topics are typically replicated across AZs, further increasing cross-AZ traffic and, consequently, cloud costs.
	\item \textbf{Operational complexity.} Operating a stateful messaging system such as Apache Kafka at large scale is inherently complex, especially when scaling clusters, performing upgrades, or recovering from broker failures. High shuffle volumes increase throughput and storage pressure on Kafka, exacerbating these operational challenges.
\end{itemize}

In this paper, we introduce \emph{BlobShuffle}, a novel approach to cost-effective repartitioning for stream processing systems. BlobShuffle leverages cloud object storage systems such as Amazon S3 as an intermediate layer for shuffling data, thereby reducing the load on Kafka brokers and mitigating cross-AZ traffic costs. Instead of sending all shuffled records through Kafka, BlobShuffle groups records into batches, stores these batches in object storage, and forwards only compact notifications via Kafka that reference the stored batches. Downstream operators then retrieve the referenced batches from object storage and extract the relevant records.
This design trades lower shuffle cost for higher shuffle latency on the order of a few seconds, which is acceptable for many high-throughput workloads such as continuous observability analytics with windowed aggregations, as we encounter in production at Dynatrace.
In contrast to recent efforts to build messaging systems natively on top of object storage~\cite{Merli2025,WarpStream,AutoMQ,Inkless,KIP-1150}, our approach is integrated purely on the client side, avoids replacing the central messaging backbone, and can therefore be adopted in existing deployments with minimal disruption.

BlobShuffle is designed to balance cost efficiency and latency through configurable batching and a distributed caching mechanism that reduces redundant object storage accesses.
We implement BlobShuffle as an add-on for Kafka Streams that requires only minimal code changes to existing applications, as illustrated in \cref{lst:example}. This design makes BlobShuffle applicable to a broad range of existing Kafka-based deployments without requiring changes to Kafka itself or to the surrounding infrastructure.
At the same time, BlobShuffle integrates tightly with Kafka Streams' mechanisms for correctness, consistency, and fault tolerance~\cite{Wang2021}: it is integrated into the framework's commit protocol to ensure that batches are durably stored and referenced before state is committed, thereby preserving Kafka Streams' at-least-once and exactly-once processing guarantees. %

Beyond a theoretical analysis of cost and latency trade-offs, we conduct a large-scale experimental evaluation in a real cloud deployment.
We systematically study how different batch sizes affect both shuffle latency and cloud costs and show that BlobShuffle can reduce shuffling costs by more than $40\times$ compared to native Kafka Streams shuffling while keeping the 95th percentile latency below 2~seconds.
Moreover, we evaluate BlobShuffle's scalability by running it with up to 48~Kafka Streams instances processing more than 2~gigabytes of uncompressed data per second, demonstrating that it can support shuffle-intensive workloads at production scale.

\paragraph{Contributions}
In summary, we make the following contributions with this paper:

\begin{itemize}
	\item We present BlobShuffle, a novel approach to cost-effective repartitioning for stream processing systems that leverages cloud object storage and demonstrate its integration as an add-on for Kafka Streams.\footnote{The source code of BlobShuffle is available at \url{https://github.com/dynatrace-research/BlobShuffle}.}
	\item We develop an analytical model that captures the cost and latency characteristics of BlobShuffle and enables reasoning about configuration trade-offs.
	\item We perform an extensive experimental evaluation in a realistic cloud deployment, quantifying BlobShuffle's cost savings, performance implications, and scalability for different batch sizes.
\end{itemize}

\begin{listing}
\begin{minted}[fontsize=\footnotesize,breaklines=true,frame=lines,framerule=\heavyrulewidth]{diff}
@@ -90,7 +90,18 @@
+BlobShuffle<MyKey, MyValue> blobShuffle = new BlobShuffle<>(
+  new S3Config(s3Client, "my-bucket-name"),
+  LocalCacheConfig.enabled(cacheSizeBytes, cacheOnWrite),
+  DistributedCacheConfig.enabled(cacheSizeBytes, cacheOnWrite, advertisedHost, port),
+  System.getenv("TOPOLOGY_ZONE"),
+  Serdes.MyKey(),
+  Serdes.MyValue(),
+  16 * 1024 * 1024, // desired batch size
+  Duration.ofSeconds(5)); // max. batch duration
 StreamsBuilder builder = new StreamsBuilder();
 builder.stream(kafkaTopicInput, Consumed.with(...))
   .flatMap((k, v) -> ...)
-  .repartition(Repartitioned.with(Serdes.MyKey(), Serdes.MyValue()))
+  .process(blobShuffle.batcherSupplier())
+  .repartition(blobShuffle.repartitioned())
+  .process(blobShuffle.debatcherSupplier())
   .aggregate(...)
   .to(kafkaTopicOutput, Produced.with(...));
 Topology topology = builder.build();
\end{minted}
\caption{Example of integrating BlobShuffle into an existing Kafka Streams application with only minimal code changes required. The original repartition step is replaced by BlobShuffle's Batcher and Debatcher operators and a repartitioning step configured by BlobShuffle.}
\label{lst:example}
\end{listing}

\paragraph{Outline}
\Cref{sec:background} introduces the necessary background on shuffling in distributed stream processing, the Kafka Streams framework, and cloud object storage.
\Cref{sec:approach} presents the design of BlobShuffle.
\Cref{sec:theory} develops an analytical cost and latency model.
\Cref{sec:evaluation} evaluates BlobShuffle in a realistic cloud environment.
\Cref{sec:beyond-kstreams} discusses how our BlobShuffle approach can be applied to other stream processing frameworks and use cases beyond Kafka Streams.
\Cref{sec:related-work} discusses related work and \cref{sec:conclusions} concludes the paper.

\section{Background}\label{sec:background}

In the following, we provide background on shuffling in distributed stream processing, the Kafka Streams framework, and cloud object storage to motivate and contextualize the design of BlobShuffle.

\subsection{Shuffling in Distributed Stream Processing}

Distributed stream processing systems and frameworks achieve scalability by partitioning input streams and processing different partitions in parallel across multiple instances. While the isolated processing of data records remains unaffected by the assignment of data portions to instances, processing logic that depends on previous records, such as aggregations, requires state management. 

To support such stateful workloads efficiently, modern stream processing frameworks adopt a design similar to the MapReduce~\cite{Dean2008} programming model. That means, before a stateful operation in the processing pipeline, a dedicated operator assigns keys to records. The stream processing framework then redistributes records over the network so that all records with the same key are routed to the same processing instance. This step is commonly referred to as \emph{shuffling} or \emph{repartitioning}. Consequently, the stateful operator can maintain per-key state locally without any need for distributed synchronization. By avoiding fine-grained state sharing and coordination across instances, shuffling enables stream processing applications to scale at significantly lower costs than architectures that rely on frequent remote access to shared state backends, such as typical Functions-as-a-Service deployments~\cite{IC2E2022}.

\subsection{The Kafka Streams Framework}

Kafka Streams~\cite{Sax2018,Wang2021} is one of the most widely used open-source stream processing frameworks and is heavily adopted in industry. %
In contrast to many other stream processing systems such as Flink or Spark, Kafka Streams is designed as a lightweight Java library that is embedded into standalone applications rather than deployed as a separate cluster. This library-style design fits well with microservice architectures, where each service owns its processing logic and state and can be independently deployed and scaled~\cite{Laigner2025,JSS2024}.

Applications built with Kafka Streams define a directed acyclic graph (DAG) of operators, called a \emph{topology}. Records are represented as key--value pairs that flow through this topology and are transformed, filtered, or aggregated by operators. For scalability, Kafka Streams requires partitioned input topics and executes the topology in parallel across multiple \emph{instances} (processes).
Each instance runs one or more threads and each thread executes one or more \emph{tasks}, where a task is the basic execution unit that processes exactly one partition of each input topic. Kafka Streams uses Kafka itself not only as the messaging backbone, but also for coordination and fault tolerance, for example through consumer groups, offset management, and changelog topics for state.

Stateful operators---such as aggregations, joins, and windowed computations---maintain per-key state in local state stores. To avoid synchronized access to shared state and to enable efficient local updates, Kafka Streams requires that all records with the same key are processed by the same task. When the upstream data distribution does not already satisfy this requirement, Kafka Streams performs a \emph{repartitioning} before the stateful operator. In practice, this is realized by splitting the topology at the shuffle boundary: the upstream sub-topology writes records to an internal intermediate Kafka topic (a \emph{repartitioning topic}), from which the downstream sub-topology consumes. This design decouples the upstream and downstream parts of the topology and leverages Kafka's durability and ordering guarantees for shuffling.

In cloud deployments across multiple availability zones (AZs), this shuffle pattern can have important cost implications. Producing records from the upstream sub-topology to a repartitioning topic may cross AZ boundaries between the producer instance and the responsible Kafka broker. In addition, Kafka replicates partitions of the repartitioning topic across brokers, which are often placed in different AZs for fault tolerance. While Kafka Streams can assign consumer tasks to brokers within the same AZ to reduce cross-AZ traffic on the consumption side, cross-AZ traffic on the producer side and for replication is difficult to avoid. As a result, shuffle-intensive Kafka Streams applications can incur substantial cross-AZ network costs and impose significant throughput and storage load on the Kafka cluster.

\subsection{Cloud Object Storage}

Cloud object storage services such as Amazon S3, Google Cloud Storage, or Azure Blob Storage provide scalable and fault-tolerant storage for large volumes of data. They expose an HTTP-based API for reading and writing immutable \emph{objects}, which are addressed by a key within a logical container (e.g., a \textit{bucket}). In contrast to direct machine-to-machine network communication, object stores are not designed for low-latency access. Requests often exhibit significantly higher latency, but the services are engineered to scale to very high throughput and capacity with strong durability guarantees.

The pricing model of cloud object storage is fundamentally different from that of direct machine-to-machine network communication. Providers charge per stored byte and per operation type (e.g., PUT, GET, LIST), where the cost of an operation usually does not depend on the size of the object within broad limits. This makes object stores attractive for coarse-grained, batched access patterns, where a small number of large objects can replace a large number of fine-grained operations.

Object storage is typically offered as a region-wide service without an explicit notion of availability zones in its interface. From the perspective of the client, an object written in one AZ can be read from another AZ without incurring additional cross-AZ data transfer charges; the provider hides replication and placement decisions internally. As a result, object storage can act as a cost-efficient, fault-tolerant foundation for data exchange across AZ boundaries, provided that the higher access latency is amortized through batching and caching.

\section{The BlobShuffle Approach}\label{sec:approach}

\begin{figure}%
	\begin{subfigure}[b]{\linewidth}%
		\includegraphics{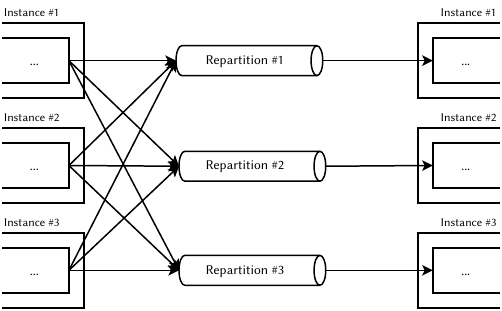}%
		\caption{Without BlobShuffle}%
		\label{fig:overview:without}%
	\end{subfigure}
	
	\vspace{0.5em}
	
	\begin{subfigure}[b]{\linewidth}%
		\includegraphics{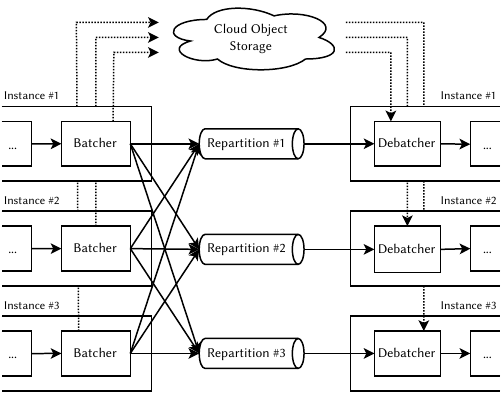}%
		\caption{With BlobShuffle}%
		\label{fig:overview:with}%
	\end{subfigure}
	\caption{Shuffling in Kafka Streams without and with BlobShuffle.}
	\label{fig:overview}
\end{figure}

The fundamental idea of our BlobShuffle approach is to batch records before the shuffling, write those batches to a cloud object storage service and then forward only notifications with a reference to those batches. When the notification is received, BlobShuffle fetches the corresponding batch from the object storage service and extracts all records in that batch that are referenced by the notification for further processing. A multi-layered cache helps to reduce latency and costs for requests to the object storage service.
\Cref{fig:overview} illustrates the high-level dataflow for shuffling with BlobShuffle in contrast to the Kafka Streams' default.
We describe both the write path (i.e., creating and storing batches of records) and the read path (i.e., retrieving batches and extracting the contained records) as well as the caching mechanism in detail.

\subsection{Write Path}

\begin{figure*}
	\centering
	\includegraphics[page=2]{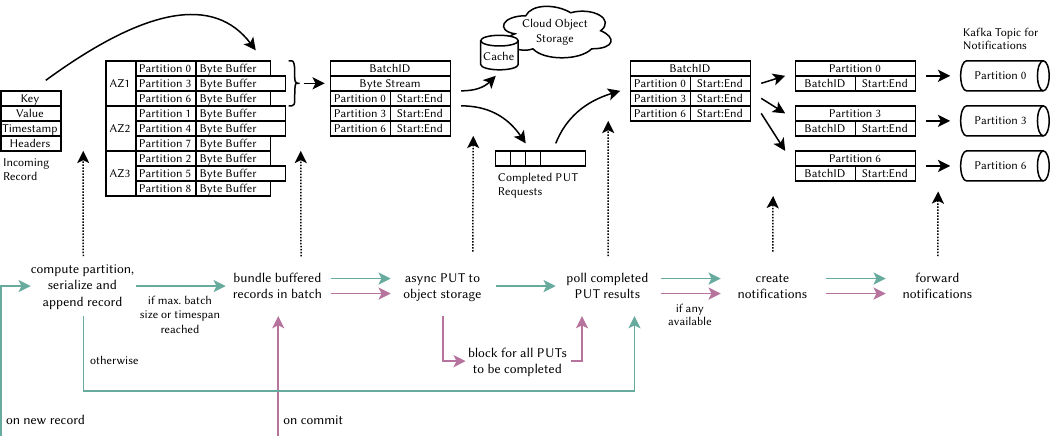}
	\caption{Overview of BlobShuffle's Batcher operator. Incoming records are grouped into batches, uploaded to the cloud object storage service, optionally cached, and corresponding notifications with references to these batches are published via Kafka.}
	\label{fig:write-path}
\end{figure*}

In a standard Kafka Streams application, records are assigned to partitions using a configurable partitioning function that maps the record key to a partition number. With BlobShuffle, this partition assignment is performed inside the Batcher operator (cf.~\cref{fig:write-path}). For each destination partition, BlobShuffle maintains an in-memory buffer that stores records in serialized form (including key, value, timestamp, and headers). Buffers for partitions that reside in the same availability zone (AZ) are additionally grouped together, allowing BlobShuffle to track the accumulated buffer size per AZ. Buffers are shared across all stream tasks executed within the same streams thread, and can optionally be shared across all threads on the same instance.

When a new record arrives, the Batcher computes its destination partition and appends the serialized record to the corresponding buffer. For each partition, the Batcher also keeps track of the AZ in which the partition is located. This information is used when forming batches: once a batch is finalized, it will always consist of the concatenated contents of all buffers belonging to partitions in the same AZ. Thus, the resulting batch is a single byte buffer composed of the per-partition byte buffers, such that records for a given partition appear sequentially within the batch.

A batch is completed when one of the following conditions is met: (i) the maximum configured batch size is reached, (ii) a maximum batching interval since the last batch finalization elapses, or (iii) Kafka Streams initiates a commit of the current processing progress. When a batch is finalized, it is assigned a unique identifier and asynchronously uploaded to the cloud object storage service. Optionally, the batch may also be inserted into a cache (see \cref{sec:blobshuffle:caching}). While the upload is in progress, the Batcher allocates fresh buffers so that subsequent records can be processed without blocking.

To decouple upload completion from the main record-processing loop, the Batcher maintains an internal queue of upload results. Once the upload of a batch finishes, an item is appended to this queue containing the batch identifier and a compact index mapping partition identifiers to their corresponding byte ranges within the batch. During normal record processing, the Batcher periodically polls this queue from the main processing thread, ensuring that the handling of asynchronous uploads does not introduce side effects in separate threads. For each partition that contributed records to the batch, the Batcher sends a notification via the repartitioning topic that contains a reference to the batch, namely the batch identifier and the byte range for that partition.

A special case occurs when Kafka Streams attempts to commit the current processing progress. In this situation, the Batcher temporarily blocks the commit until all outstanding asynchronous uploads have completed, their results have been processed from the internal queue, and all corresponding notifications have been sent. This ensures that state is only committed once the derived batches have been durably stored in object storage and the references to these batches have been published to Kafka. Consequently, failures during batch upload or finalization cause the commit to fail, leading Kafka Streams to roll back to the latest committed state and preserving its at-least-once and exactly-once processing guarantees~\cite{Wang2021}. Note that it may happen that a batch is successfully uploaded to object storage, but no corresponding notification is ever sent (e.g., due to a failure before commit). Such orphaned batches are harmless, as they remain unreachable: BlobShuffle and downstream operators never reference them and the object storage service is treated as an append-only, garbage-tolerant storage layer.

\subsection{Read Path}

\begin{figure*}
	\centering
	\includegraphics[page=2]{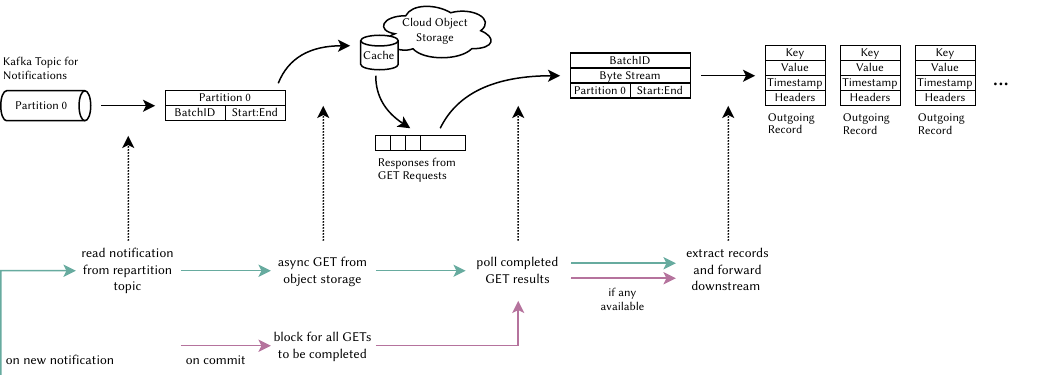}
	\caption{Overview of BlobShuffle's Debatcher operator. Received notifications reference batches, which are retrieved from object storage or the cache, and the contained records are extracted and forwarded one by one to the downstream operator.}
	\label{fig:read-path}
\end{figure*}

BlobShuffle uses a Debatcher operator that processes the notifications emitted by the Batcher (cf.~\cref{fig:read-path}). The Debatcher reads these notifications from the repartitioning topic. Each notification specifies a batch identifier and a byte range for a particular partition. Based on this information, the Debatcher issues an asynchronous read request to the cloud object storage service (or to the cache, see \cref{sec:blobshuffle:caching}) to retrieve the corresponding batch or the relevant subset of the batch (sub-batch).

Similar to the write path, the results of these asynchronous requests are decoupled from the main processing loop. The Debatcher maintains an internal queue to which the returned batch data are appended once the asynchronous reads complete. The normal processing loop of the Debatcher periodically polls this queue. For each received batch or sub-batch, it deserializes the contained records and forwards them, one by one, to the downstream operators in Kafka Streams.

When Kafka Streams initiates a commit, the Debatcher must again ensure that no in-flight asynchronous operations remain. It therefore blocks the commit until all outstanding read requests have completed and all records contained in the corresponding batches have been fully processed. This coordination mirrors the behavior on the write path and preserves Kafka Streams' at-least-once and exactly-once processing guarantees.

BlobShuffle does not explicitly delete batches from object storage after they have been read. Instead, batches are removed automatically after a configurable retention period. This approach is similar to Kafka's log retention and is acceptable because storing batches for a limited time (e.g., few hours) incurs only modest storage costs.

\subsection{Caching}\label{sec:blobshuffle:caching}

\begin{figure*}%
	\subcaptionbox{With local cache\label{fig:cache:wolocal}}{%
		\includegraphics[page=2]{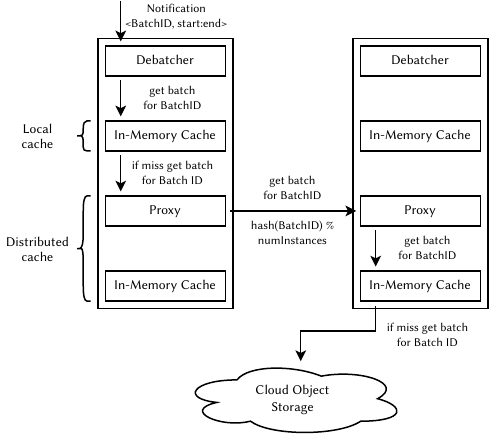}%
	}\hfill%
	\subcaptionbox{Without local cache\label{fig:cache:wlocal}}{%
		\includegraphics[page=2]{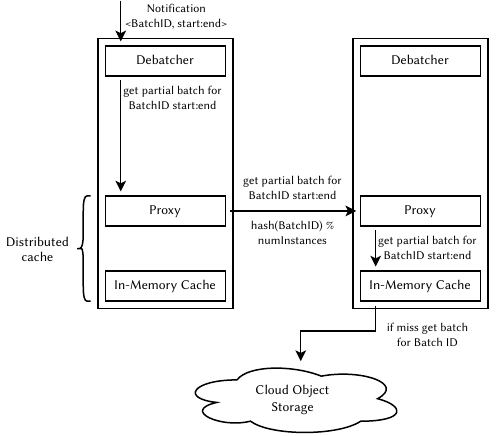}%
	}
	\caption{Caching mechanisms in BlobShuffle's read path with and without an additional local cache layer.}
	\label{fig:cache}
\end{figure*}

All access to the object storage service is routed through a potentially multi-layered caching mechanism. In our current implementation, BlobShuffle employs a combination of a distributed in-memory Least Recently Used (LRU) cache and an optional local in-memory LRU cache as illustrated in \cref{fig:cache}. Additional cache layers, such as disk-based caches, could be integrated depending on the use case and its deployment constraints.

\subsubsection*{Distributed cache}

The distributed cache is organized per availability zone. All stream processing instances within the same AZ form a cache cluster, where each member is responsible for maintaining cached entries for a subset of batches. All read and write requests for batches are routed through the corresponding cluster member. For write requests, the cluster member forwards the batch to the object storage service and, optionally, stores it in the distributed cache. For read requests, it first checks whether the requested batch is present in the cache and if not, it downloads the batch from object storage and inserts it into the cache.

To avoid redundant downloads, the distributed cache coordinates concurrent requests: if multiple read requests for the same batch arrive while the batch is still being downloaded, subsequent requests are blocked until the initial download completes and the batch becomes available in the cache. In this way, the distributed cache ensures that a given batch is downloaded from object storage at most once per AZ (unless the corresponding cache entry has expired), which reduces object storage access costs and lowers latency for repeated access to the same batch. Depending on the deployment and batch size, stream processing instances may either fetch complete batches, including records for all partitions, or sub-batches restricted to a single partition. Fetching complete batches is required to make effective use of the local cache, as it allows subsequent accesses to different partitions within the same batch to be served from the cache.

\subsubsection*{Local cache}

In addition to the distributed cache, BlobShuffle can employ an optional local cache on each stream processing instance to further improve resource efficiency. When enabled, all read and write operations first go through the local cache. For read requests, the local cache checks whether the batch is already stored locally and only forwards the request to the distributed cache if the batch is missing. This avoids repeated remote lookups for batches that are frequently accessed by the same instance.

The local cache is particularly beneficial when a large number of partitions are processed by a single Kafka Streams instance. In such scenarios, multiple operators on the same instance might otherwise trigger separate requests for the same batch via the distributed cache. By serving these accesses from a local LRU cache, BlobShuffle reduces network traffic within the AZ and improves end-to-end latency for batch retrieval. %

\section{Analytical Cost and Latency Model}\label{sec:theory}

To reason about the cost and latency characteristics of BlobShuffle, we introduce a simple analytical model under steady-state conditions.
We first define the relevant parameters and then derive the expected batching rate, object storage operation rates, and shuffle latency.

\subsection{Model Parameters}

We assume a deployment with the following parameters:

\begin{itemize}
	\item $N_{\mathrm{inst}}$: number of stream processing instances.
	\item $N_{\mathrm{az}}$: number of availability zones (AZs).
	\item $\lambda$: total input rate in records per second (aggregated across all instances).
	\item $s_{\mathrm{rec}}$: average record size in bytes.
	\item $S_{\mathrm{batch}}$: target batch size in bytes.
	\item $T_{\mathrm{put}}$: latency of a PUT request to the object storage service (from issuing the request until the batch is durably stored).
	\item $T_{\mathrm{get}}$: latency of a GET request from the object storage service (from issuing the request until the batch is available at the consumer).
\end{itemize}

We assume that records are evenly distributed across instances and AZs and that batching is performed independently per AZ over all instances in that AZ.
Under these assumptions, each stream processing instance observes an average input rate of
\[
\lambda_{\mathrm{inst}} = \frac{\lambda}{N_{\mathrm{inst}}} \quad \text{[records/s per instance].}
\]
The corresponding byte rate per instance is
\[
b_{\mathrm{inst}} = \lambda_{\mathrm{inst}} \cdot s_{\mathrm{rec}} = \frac{\lambda \, s_{\mathrm{rec}}}{N_{\mathrm{inst}}} \quad \text{[bytes/s per instance].}
\]

\subsection{Batch Formation and Request Rates}

\paragraph{Time to fill a batch.}
Since batches are formed per target AZ, the expected time to fill a batch of size $S_{\mathrm{batch}}$ on a given stream processing instance is
\[
T_{\mathrm{batch}} = \frac{S_{\mathrm{batch}} \, N_{\mathrm{az}}}{b_{\mathrm{inst}}}
= \frac{S_{\mathrm{batch}} \, N_{\mathrm{az}} \, N_{\mathrm{inst}}}{\lambda \, s_{\mathrm{rec}}}
\quad \text{[s per batch].}
\]

\paragraph{Batches per second.}
In steady state, each stream processing instance completes on average
\[
\mu_{\mathrm{batch,inst}} = \frac{N_{\mathrm{az}}}{T_{\mathrm{batch}}}
= \frac{\lambda \, s_{\mathrm{rec}}}{S_{\mathrm{batch}} \, N_{\mathrm{inst}}}
\quad \text{[batches/s per instance].}
\]
Across all AZs, the system therefore produces
\[
\mu_{\mathrm{batch}} = N_{\mathrm{inst}} \cdot \mu_{\mathrm{batch,inst}}
= \frac{\lambda \, s_{\mathrm{rec}}}{S_{\mathrm{batch}}}
\quad \text{[batches/s].}
\]

\paragraph{Object storage request rates.}
Each completed batch triggers exactly one PUT request, so the PUT request rate is
\[
\mu_{\mathrm{put}} = \mu_{\mathrm{batch}} 
= \frac{\lambda \, s_{\mathrm{rec}}}{S_{\mathrm{batch}}}
\quad \text{[PUT/s].}
\]
Assuming that each batch is consumed in exactly one AZ, and that consumption in the producing AZ can be served from the cache while consumption in any of the other $N_{\mathrm{az}} - 1$ AZs requires fetching the batch from the object storage service (after which it is cached for subsequent requests to the same batch), the GET request rate is
\[
\mu_{\mathrm{get}} = \mu_{\mathrm{batch}} \cdot \frac{N_{\mathrm{az}} - 1}{N_{\mathrm{az}}}
= \frac{\lambda \, s_{\mathrm{rec}}}{S_{\mathrm{batch}}} \cdot \frac{N_{\mathrm{az}} - 1}{N_{\mathrm{az}}}
\quad \text{[GET/s].}
\]

\subsection{Shuffle Latency}

We focus on the latency introduced by the shuffle for an individual record, measured from the time it enters the Batcher until it is emitted by the Debatcher.
For a record that arrives immediately after a new batch has been started, the worst-case waiting time until the batch is full is $T_{\mathrm{batch}}$.
Afterward, the batch is uploaded (PUT) and later retrieved (GET) in the consuming AZ.
A simple upper bound on the shuffle latency is therefore
\[
T_{\mathrm{shuffle}}^{\mathrm{max}} = T_{\mathrm{batch}} + T_{\mathrm{put}} + T_{\mathrm{get}}.
\]

Additional latency introduced by transmitting notifications and accessing the cache is negligible compared to object storage operations.
In practice, latency is often smaller, depending on when within the interval $T_{\mathrm{batch}}$ a record arrives, and for records that are shuffled to the same AZ as no GET request is required for those.
For a constant input rate $\lambda$ and number of AZs $N_{\mathrm{az}}$, the bound $T_{\mathrm{batch}} + T_{\mathrm{put}} + T_{\mathrm{get}}$ highlights the dependence of shuffle latency on the batch size $S_{\mathrm{batch}}$ and on the object storage latencies $T_{\mathrm{put}}$ and $T_{\mathrm{get}}$, which typically also depend on $S_{\mathrm{batch}}$ through request processing and data transfer time~\cite{Durner2023}.
$T_{\mathrm{put}}$ and $T_{\mathrm{get}}$ are not constants, but follow a typically long-tail distribution.
Due to Kafka Streams' periodic commits, durability is not impacted by higher latency. %

\section{Experimental Evaluation}\label{sec:evaluation}

In addition to the theoretical analysis in \cref{sec:theory}, we experimentally evaluate the performance and cost characteristics of BlobShuffle and its Kafka Streams implementation in a cloud-native deployment.
Specifically, we investigate:

\begin{enumerate}
	\item How does our design influence the latency distribution?
	\item How does the batch size affect latency and cost?
	\item How does the partition count influence coordination overhead and throughput?
	\item How well does BlobShuffle scale with more instances?
\end{enumerate}

\subsection{Methodology and Setup}\label{sec:evaluation:method}

In addition to the following description, we provide the exact evaluation setup and raw results as supplemental material~\cite{ReplicationPackage}, allowing other researchers to repeat and extend our work.

\subsubsection{Benchmark application}

To evaluate BlobShuffle, we integrate it into a Kafka Streams application.
Instead of using one of the available end-to-end benchmarks, we construct a minimal benchmark application focusing on a single shuffle step so that observed latency and cost effects can be attributed directly to BlobShuffle and its configuration rather than to unrelated application-specific processing.
Such a shuffle operation is part of all benchmarks that include stateful aggregations~\cite{Wang2024}.
The application implements a linear Kafka Streams topology consisting of the following steps:
\begin{enumerate}[label=(\roman*)]
	\item Reading records from an \emph{input} Kafka topic. Each record contains a value with random byte content and no key.
	\item Assigning a key to each record based on the first 8~bytes of the record value. This key is later used to determine the target partition of this record.
	\item Overwriting the last bytes of the record value with the current processing timestamp.
	\item Performing the BlobShuffle-based shuffle, i.e., batching, storing batches in object storage and caches, exchanging notifications, and downstream debatching.
	\item Measuring the shuffle latency for each record by comparing the previously written timestamp with the current processing time and exporting this latency to a monitoring system.
	\item Finally, writing the record to an \emph{output} Kafka topic.
\end{enumerate}

The random data in the input topic is generated independently of Kafka Streams using the ShuffleBench load generator~\cite{ICPE2024}.

\subsubsection{Cloud-native deployment}

We conduct our experimental evaluation in a Kubernetes cluster managed by the Elastic Kubernetes Service (EKS) of Amazon Web Services (AWS) in the \textit{us-east-1} region.
The cluster consists of three groups of nodes:
Depending on the experiment, up to 18 \textit{m6i.xlarge} EC2 instance nodes run the load generators as well as additional benchmarking infrastructure.
9 \textit{m6in.2xlarge} nodes run the Kafka brokers, distributed across three availability zones, such that each AZ runs three brokers.
Depending on the experiment, up to 12 or 24 \textit{r6in.xlarge} nodes run the Kafka Streams benchmark application, again distributed across three AZs, resulting in either up to 4 or 8 nodes per AZ.
On each application node, we run two instances of the Kafka Streams application, resulting in a total of up to 24 or 48 instances.
As the object storage service, we use Amazon S3 in the same \textit{us-east-1} region.

The relatively large Kafka cluster is primarily required to handle the high load on the input and output topics of the benchmark infrastructure rather than to support the BlobShuffle shuffle itself.
Similarly, we use network-optimized EC2 instance types for the Kafka Streams application because, in our minimal benchmark, network overhead dominates compute and memory; a real-world stream processing application with more complex processing logic would likely be compute or memory-bound, allowing for more cost-effective EC2 instance types.
Except for configuring AZ awareness and, in some experiments, increasing selected timeouts for large deployments, we use Kafka Streams with its default configuration.

We utilize the cloud-native benchmarking framework Theodolite~\cite{EMSE2022} to automate benchmark execution in Kubernetes.
Theodolite allows us to define experiments declaratively, handles the setup and teardown of all involved software components, and collects the measurement data used in our evaluation.

\begin{figure*}%
	\hspace{0.319447in}%
	\subcaptionbox{record shuffle latency\label{fig:latency:total}}{%
		\hspace{-0.319447in}%
		\includegraphics{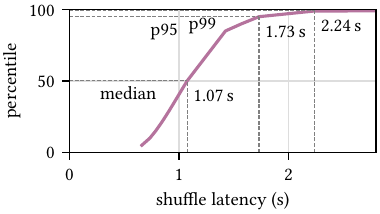}%
	}\hfill%
	\subcaptionbox{batch PUT latency\label{fig:latency:put}}{%
		\includegraphics{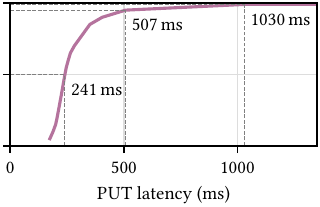}%
	}\hfill%
	\subcaptionbox{batch GET latency\label{fig:latency:get}}{%
		\includegraphics{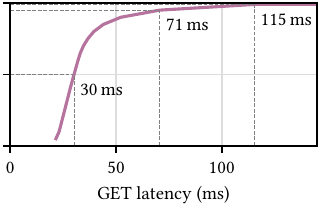}%
	}%
	\caption{Latency distributions for BlobShuffle (24~Kafka Streams instances on 12~nodes with 16\,MiB batches).}%
	\label{fig:latency}
\end{figure*}

\subsubsection{Workload and BlobShuffle configuration}

We configure the load generator to produce records of size 1\,KiB.
Each load generator instance can emit up to 180\,000 records per second, resulting in 3.24~million records per second across all instances.
Unless indicated otherwise, BlobShuffle is configured with a target batch size of 16\,MiB.
We enable the distributed cache with a capacity of 4\,GiB per application instance and disable the local cache.

We scale the number of partitions on all Kafka topics proportional to the number of Kafka Streams application instances and, unless stated otherwise, set it to $9\times$ the number of application instances.
Given three Kafka brokers per availability zone and a uniform partition assignment, this choice ensures that each Kafka Streams instance consumes data from exactly three partitions on each of the three brokers in its AZ.
This setup ensures that load is evenly distributed across brokers and application instances.

\subsubsection{Metrics and measurement methods}

In line with the findings of our previous study~\cite{FSE2025}, which quantified performance variability of stream processing systems in the cloud, we execute each experiment three times.
As the min--max range across repetitions is consistently very low, we report the median of the three runs throughout the evaluation.\footnote{See our supplemental material~\cite{ReplicationPackage} for the complete dataset.}
We run each experiment for 15~minutes, while considering the first 6~minutes as a warm-up period, after which our benchmark application operates in steady state for all configurations.
During the remaining time, we continuously sample performance and system data from the benchmarked application and derive the following metrics.

\paragraph{Throughput}
We measure throughput according to the \textit{ad-hoc throughput} method as presented and evaluated in our previous work~\cite{ICPE2024}.
Essentially, this means we generate data at a rate higher than the system can process and continuously measure the rate of actually processed records.%
\footnote{Ad-hoc throughput tends to overestimate the \textit{sustainable throughput} that could be expected in a real-world production deployment~\cite{ICPE2024}, because it ensures that internal batches (e.g., in Kafka) can be fully utilized when fetching data. However, conducting reliable sustainable throughput measurements, requires a large number of carefully controlled experiments~\cite{LTB2021,EMSE2022,ICPE2024}, which would significantly increase evaluation time and cloud costs. For our purposes, we consider the ad-hoc throughput method acceptable, as the overestimation is systematic across all configurations and our focus is on benchmarking BlobShuffle rather than Kafka Streams itself.}

\paragraph{Latency.}
As described above, we measure per-record latency as the elapsed time between writing the current timestamp immediately before the record enters the BlobShuffle Batcher and reading that timestamp immediately after the record leaves the Debatcher.

\paragraph{Costs.}
We compute cloud costs for all relevant services based on provider list prices, specifically using Amazon Web Services pricing for the \textit{us-east-1} region at the time of writing.

\subsection{Evaluation of Latency Distribution}
\label{sec:evaluation:latency-dist}

We first briefly characterize the shuffle latency distribution of BlobShuffle.
\Cref{fig:latency} shows CDFs for the per-record shuffle latency as well as for batch uploads and downloads (PUT and GET requests to Amazon S3) for 24~Kafka Streams instances with 16\,MiB batches.

The shuffle latency distribution (\cref{fig:latency:total}) exhibits a pronounced long tail, increasing from a median of 1.07\,s to 1.73\,s at the 95th percentile and 2.24\,s at the 99th percentile.
This is mostly explained by the long-tailed latencies of object storage operations: both PUT (\cref{fig:latency:put}) and GET (\cref{fig:latency:get}) latencies approximately double from the median to p95 and again from p95 to p99.
Notably, PUT requests are about 7--9$\times$ slower than GET requests and, therefore, have a much stronger impact on the overall shuffle latency.
We observe similar long-tailed distributions for our other evaluated configurations.

\begin{resultsbox}{Finding:}
	BlobShuffle's shuffle latency is long-tailed and largely driven by the long-tailed latency of object storage PUT operations.
\end{resultsbox}

\subsection{Evaluation of Batch Sizes}

As mentioned before, the batch size is a critical parameter that influences both performance and cost of BlobShuffle. Larger batches amortize the overhead of object storage operations across more data, reducing the number of PUT and GET requests per byte and, thus, lowering costs.
At the same time, larger batches take longer to fill and therefore increase end-to-end latency.
Conversely, smaller batches reduce waiting time in the Batcher but increase the number of object storage operations and associated costs.

In this experiment, we evaluate different target batch sizes ranging from 1\,MiB to 128\,MiB.
Note that the configured batch size is a target, actual batch sizes can be smaller when a batch is finalized by a commit event.
We keep the maximum batch duration sufficiently large, ensuring that batches are typically not closed due to timing.
For this evaluation, we use a cluster with 12 nodes running 24 Kafka Streams application instances (pods), as described in \cref{sec:evaluation:method}.

\begin{figure*}%
	\subcaptionbox{total throughput\label{fig:batchsize:throughput}}{%
		\includegraphics{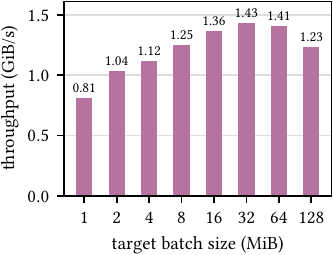}%
	}\hfill%
	\subcaptionbox{throughput per node\label{fig:batchsize:throughput-node}}{%
		\includegraphics{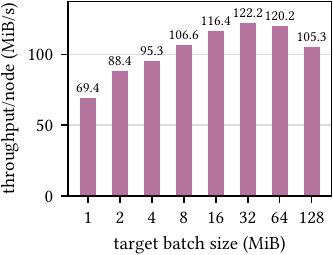}%
	}\hfill%
	\subcaptionbox{latency\label{fig:batchsize:latency}}{%
		\includegraphics{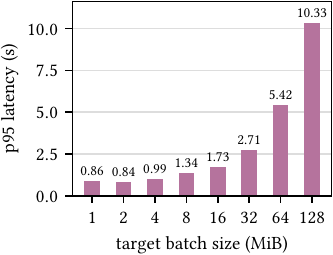}%
	}
	
	\vspace{0.5em}
	
	\subcaptionbox{PUT requests\label{fig:batchsize:puts}}{%
		\includegraphics{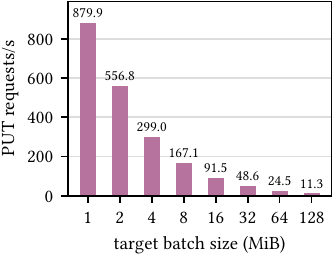}%
	}\hfill%
	\subcaptionbox{GET requests\label{fig:batchsize:gets}}{%
		\includegraphics{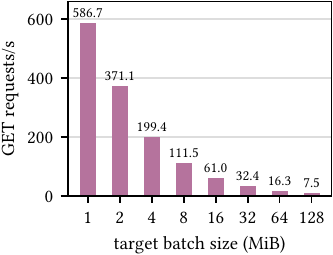}%
	}\hfill%
	\subcaptionbox{GET/PUT requests\label{fig:batchsize:gets-per-puts}}{%
		\includegraphics{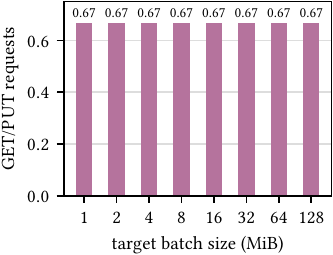}%
	}
	
	\vspace{0.5em}
	
	\subcaptionbox{actual batch size\label{fig:batchsize:actual-ratio}}{%
		\includegraphics{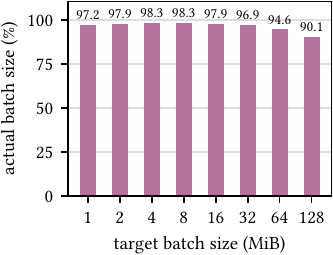}%
	}\hfill%
	\subcaptionbox{normalized S3 costs\label{fig:batchsize:s3-costs}}{%
		\includegraphics{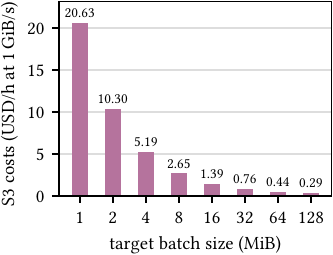}%
	}\hfill%
	\subcaptionbox{normalized EC2 costs\label{fig:batchsize:ec2-costs}}{%
		\includegraphics{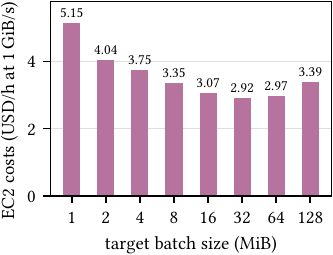}%
	}
	\caption{Impact of the target batch size on performance and costs for BlobShuffle (24~Kafka Streams instances on 12~nodes).}
	\label{fig:batchsize}
\end{figure*}

\Cref{fig:batchsize} shows various metrics collected throughout the experiments for different target batch sizes.
Throughput increases with larger batches and peaks around a target batch size of 32\,MiB with a throughput of 1.43\,GiB/s for the whole cluster (\cref{fig:batchsize:throughput}), which corresponds to 61.1\,MiB/s per pod or 122.2\,MiB/s per node (\cref{fig:batchsize:throughput-node}). For larger batches, throughput declines again.

As expected, shuffle latency grows with batch size, as illustrated for the 95th percentile (\cref{fig:batchsize:latency}).
Noteworthy, however, is that for batch sizes up to roughly 32\,MiB the latency rises less than proportional to the increase in batch size.

The average number of PUT (\cref{fig:batchsize:puts}) and GET (\cref{fig:batchsize:gets}) requests to Amazon S3 decreases markedly with increasing batch size. This is in line with the theoretical analysis in \cref{sec:theory} and the observed increase in throughput for larger batches.
It is also worth highlighting that the ratio between PUT and GET requests remains almost exactly $2{:}3$ across all configurations (\cref{fig:batchsize:gets-per-puts}), again matching the analytical model in \cref{sec:theory} and indicating that subsequent reads to the same batch are effectively served from the cache.

The average actual batch size closely tracks the configured target up to 32\,MiB (about 97--98\% of the target) and diverges slightly for larger targets (around 90\% for 128\,MiB), reflecting more frequent early finalization due to commits (\cref{fig:batchsize:actual-ratio}).

Directly related to the number of PUT and GET requests to S3 are the associated object storage costs. When we normalize S3 costs by processed data volume with one hour batch retention (\cref{fig:batchsize:s3-costs}), we observe the expected inverse relationship between batch size and cost:
assuming a constant throughput of 1\,GiB/s, the hourly S3 cost decreases from 20.63\,USD/h for 1\,MiB batches to 0.29\,USD/h for 128\,MiB batches.
In addition to object storage costs, infrastructure costs must also be considered. To estimate these, we normalize the hourly costs of the EC2 instances running the Kafka Streams application by the achieved throughput for each target batch size (\cref{fig:batchsize:ec2-costs}).
In line with the throughput results, infrastructure costs at a fixed processing rate of 1\,GiB/s are minimized at around 3.00\,USD/h for batch sizes in the range of 32--64\,MiB.
For larger batches, costs increase again because the achieved throughput decreases.

\begin{figure}
	\centering
	\includegraphics{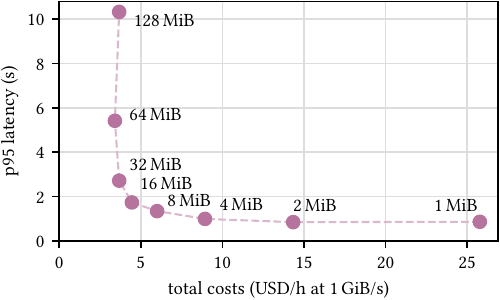}
	\caption{Latency of shuffling via BlobShuffle for different target batch sizes in relation to operating costs.}
	\label{fig:batchsize:cost-latency}
\end{figure}

\Cref{fig:batchsize:cost-latency} correlates the p95 shuffle latency with the combined costs of object storage and compute for different batch sizes at a normalized processing rate of 1\,GiB/s.
Batch sizes below 4--8\,MiB offer only limited latency improvements compared to larger batches, but incur significantly higher costs due to the increased number of object storage operations and lower throughput.
Conversely, batch sizes above 32\,MiB lead to substantially higher latency while yielding only marginal additional cost savings.
Since BlobShuffle exposes the batch size as a configurable parameter, engineers can choose a setting that best matches their latency--cost trade-off.
Our results show that for most use cases, batch sizes in the range of 8--32\,MiB provide a favorable balance between cost efficiency and latency.
For instance, 16\,MiB batches result in costs of 4.46\,USD/h at a p95 shuffle latency of 1.73\,s.
For reference, direct shuffling via Kafka would already incur 192\,USD/h in cross-AZ network charges for the same workload (assuming repartition topics replicated across three AZs to provide durability and availability comparable to BlobShuffle), resulting in overall costs well beyond a $40\times$ increase compared to BlobShuffle, although with substantially lower shuffle latency.

\begin{resultsbox}{Finding:}
	BlobShuffle enables substantial cost savings over native Kafka Streams shuffling at the expense of higher latency.
	For use cases that can tolerate up to 2~seconds latency, batch sizes in the range of 8--32\,MiB reduce shuffling costs by more than $40\times$ compared to direct Kafka-based shuffling.
\end{resultsbox}

\subsection{Evaluation of Partition Count}

\begin{figure*}%
	\subcaptionbox{throughput\label{fig:partitions:throughput}}{%
		\includegraphics{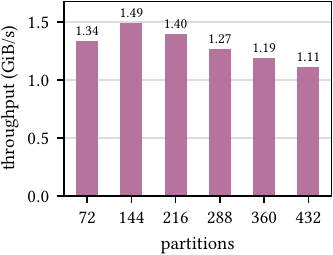}%
	}\hfill%
	\subcaptionbox{latency\label{fig:partitions:latency}}{%
		\includegraphics{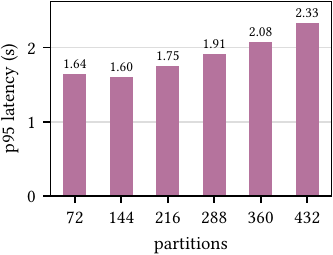}%
	}\hfill%
	\subcaptionbox{cache reads (= notifications sent)\label{fig:partitions:cache-reads}}{%
		\includegraphics{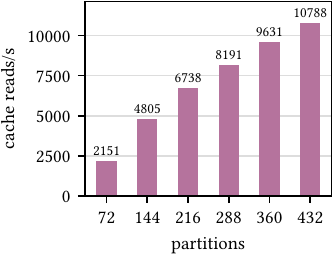}%
	}%
	\caption{Impact of scaling the number of partitions on performance for BlobShuffle (12~nodes, 16\,MiB batches).}%
	\label{fig:partitions}%
\end{figure*}

Scaling a stream processing framework such as Kafka Streams is bounded by the source data parallelism, here the number of Kafka partitions.
Before scaling the Kafka Streams deployment, we therefore evaluate how well BlobShuffle performs with more partitions.
The intuition is that more partitions increase input parallelism, but also raise BlobShuffle's coordination overhead: for a fixed batch size, each batch consists of smaller per-partition portions, which increases the number of per-partition notifications and, thus, corresponding accesses to the distributed cache, which may reduce overall throughput and increase latency.
Based on the previous results, we fix the target batch size to 16\,MiB for all experiments in this section, keep the Kafka Streams cluster fixed at 24~instances on 12~nodes and vary only the number of partitions, scaling it from $3\times$ to $18\times$ the number of instances (72 to 432 partitions in total).

\Cref{fig:partitions} summarizes metrics observed for different partition counts.
As expected, throughput decreases with more partitions (\cref{fig:partitions:throughput}), while the shuffle latency increases (\cref{fig:partitions:latency}).
This trend correlates with the rising number of batch notifications sent and the corresponding rate of reads to the distributed cache, which are effectively equivalent for BlobShuffle (\cref{fig:partitions:cache-reads}).
Despite the increase in notifications and cache reads, the impact on throughput and, consequently, on cost remains moderate: increasing the partition count by a factor of three reduces throughput by only about 26\%.
These results suggest that running with a higher, fixed number of partitions can be practical when flexible scaling of the application is required, as the throughput penalty remains limited.

\begin{resultsbox}{Finding:}
	While increasing the number of partitions leads to more per-batch notifications and corresponding reads from the cache, the resulting throughput degradation remains moderate.
\end{resultsbox}

\subsection{Evaluation of Scalability}

\begin{figure*}%
	\subcaptionbox{throughput\label{fig:scalability:throughput}}{%
		\includegraphics{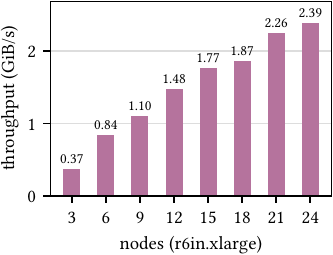}%
	}\hfill%
	\subcaptionbox{throughput per node \label{fig:scalability:throughput-instance}}{%
		\includegraphics{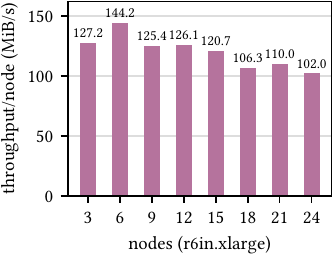}%
	}\hfill%
	\subcaptionbox{latency\label{fig:scalability:latency}}{%
		\includegraphics{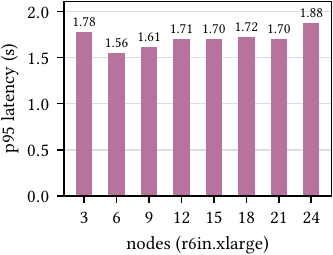}%
	}%
	\caption{Impact of scaling the cluster size on performance for BlobShuffle (16\,MiB batches).}%
	\label{fig:scalability}%
\end{figure*}

BlobShuffle targets scenarios in which the volume of data to be shuffled is so large that cross-AZ network costs dominate or where operating a message broker at very high throughput becomes an operational challenge.
We therefore evaluate the scalability of BlobShuffle by studying how performance and costs evolve with an increasing number of Kafka Streams instances.

We vary the size of the Kafka Streams application cluster from 1 to 8 nodes per availability zone, resulting in a total of 3 to 24 nodes and 6 to 48 Kafka Streams instances (Kubernetes pods).
We scale the number of partitions on all Kafka topics proportionally to the number of Kafka Streams instances.
This reflects common practice of scaling Kafka and consuming applications proportionally and avoids favoring specific cluster sizes; a fixed partition count would otherwise advantage certain cluster configurations.
In contrast to the batch size evaluation, we scale the number of partitions by a factor of 6 rather than 9 relative to the number of Kafka Streams instances, because with the higher factor the resulting total number of partitions (up to 432) caused initial Kafka Streams rebalances of several minutes for the largest cluster sizes.

\Cref{fig:scalability} shows various metrics collected throughout the experiments for different cluster sizes.
Although the throughput gains are not perfectly linear, they show a consistent increase, rising from about 0.37\,GiB/s with 3~nodes to 2.39\,GiB/s with 24~nodes (\cref{fig:scalability:throughput}).
These results suggest that no scalability limit is observed within the evaluated range.
The throughput per node decreases slightly as the number of nodes grows (\cref{fig:scalability:throughput-instance}): 6~nodes (12~pods) achieve the highest per-node throughput of 144.2\,MiB/s, whereas 24 nodes (48 pods) reach 102.0\,MiB/s per node.
Notable outliers are the configurations with 9 and 18 nodes (18 and 36 pods), which show lower per-node throughput because, in these cases, each Kafka Streams instance is effectively connected to only one Kafka broker.\footnote{Partitions are, roughly speaking, assigned in a round-robin fashion to brokers and consumers. If the number of consumers is a multiple of the number of brokers (as for the case of 18 and 36~Kafka Streams instances), each consumer reads from only one broker. Since consumer fetch requests are issued sequentially per broker, this reduces parallelism and thus throughput. This effect already occurs when reading from the input topic and is independent of BlobShuffle.}

A similar but inverse trend can be observed for latency: the 95th percentile shuffle latency increases slightly from 1.56\,s at 6~nodes to 1.88\,s at 24~nodes (\cref{fig:scalability:latency}).
This behavior is consistent with our theoretical analysis in \cref{sec:theory}, which links lower per-instance throughput to longer times needed to fill a batch.
Moreover, when we compare the per-instance throughput of the 12-instance/288-partition configuration (from the previous experiments) with the per-instance throughput of the 24-instance/288-partition configuration (from this evaluation), we observe almost identical values (108.5\,MiB/s vs. 102.0\,MiB/s).
This indicates that, for a fixed partition count, BlobShuffle scales essentially linearly with the number of instances in the evaluated range.

\begin{resultsbox}{Finding:}
	BlobShuffle scales near-linearly to at least 24 nodes (48 Kafka Streams instances), processing more than 2\,GiB/s without hitting a scalability limit.
	This is consistent with the partition-count results showing that the coherency overhead of the $n$:$n$ shuffling pattern is effectively bounded by the partition count.
\end{resultsbox}

\section{Beyond Shuffling in Kafka Streams}\label{sec:beyond-kstreams}

Although we present and evaluate BlobShuffle for repartitioning in Kafka Streams, the underlying approach is more broadly applicable.

\paragraph{Asynchronous messaging between services}
BlobShuffle can be used for general asynchronous message passing between services, applications, or systems, including publish–subscribe scenarios. In this setting, the BlobShuffle Batcher and Debatcher components are integrated into the client-side publisher and subscriber logic. The exchange of notifications that reference batches in object storage is not limited to Kafka. These notifications can also be transmitted via other messaging systems or stored and retrieved through databases. For additional decoupling, the distributed cache can be implemented as a standalone service shared by multiple clients.

\paragraph{Shuffling in other stream processing frameworks}
BlobShuffle can also support shuffling in other distributed stream processing frameworks such as Apache Flink or Apache Spark. These systems do not rely on a central messaging system like Kafka for shuffling, but instead exchange records directly between processing instances. While this design avoids cross-AZ costs for topic replication, it still incurs cross-AZ costs for direct communication between instances in different AZs. BlobShuffle remains beneficial in such scenarios: the overall design is similar, but notifications are exchanged via the framework's native shuffle mechanism (i.e., directly between instances) rather than via a separate messaging system, while batches are stored and retrieved through cloud object storage and the caching layer. 
In particular, Spark and its Structured Streaming engine appear well suited for BlobShuffle's batching approach due to their (micro-)batch processing model and comparatively high network utilization during shuffles~\cite{DEBS2024}.

\section{Related Work}\label{sec:related-work}

BlobShuffle relates to work on (i) efficient shuffling mechanisms in distributed data processing, (ii) data management systems that use cloud object storage as a primary storage layer, and (iii) messaging systems that are natively built on top of object storage. We %
discuss each area and highlight how BlobShuffle relates to these approaches.

\paragraph{Efficient shuffling in distributed processing.}
A long line of work optimizes shuffle performance for MapReduce-like batch processing jobs.
Riffle~\cite{Zhang2018} reduces small-fragment overheads in Facebook's data analytics by coalescing many tiny shuffle outputs into large, sequential blocks, significantly reducing disk seeks and improving job runtimes.
Magnet~\cite{Shen2020} introduces a push-based shuffle service at LinkedIn, where map tasks proactively send data to remote merge services that co-locate data with future reducers, reducing network and I/O overhead.
Cherry~\cite{Nikitas2021} revisits shuffle for containerized and serverless environments and proposes a stateless remote shuffle layer with task-aware caching to handle high churn.
Exoshuffle~\cite{Luan2023} decouples the shuffle control and data planes by expressing shuffle logic in a separate execution framework, enabling flexible experimentation with different shuffle strategies.
In contrast to those works, BlobShuffle focuses on optimizing costs and operational complexity for shuffling in stream processing systems.

In prior work, we introduced ShuffleBench~\cite{ICPE2024}, a benchmark designed to evaluate state-of-the-art stream processing frameworks for shuffle-intensive workloads.
We used ShuffleBench to benchmark Kafka Streams, Flink, Spark Structured Streaming, and Hazelcast for throughput and latency~\cite{ICPE2024} as well as fault-tolerance~\cite{DEBS2024}.

\paragraph{Data management systems on object storage.}
Using cloud object storage as a primary data substrate for data management systems has been studied for more than a decade~\cite{Brantner2008}, and several commercial systems now rely on object storage for durability and scalability~\cite{Dageville2016,Verbitski2017,Armbrust2020}.
Recent research, for example, investigated how to exploit cloud object storage for high-performance analytical workloads~\cite{Durner2023} or analyze cost-performance trade-offs for transactional workloads on object storage~\cite{Haubenschild2025}.
Compared to alternative storage backends, object storage provides the lowest storage cost along with high durability at the expense of much higher access latency~\cite{Durner2023,Haubenschild2025}.

Recent work on disaggregated state management in Apache Flink~2.0~\cite{Mei2025} decouples compute and state by moving operator state to remote storage, thereby improving elasticity and fault tolerance for stateful stream processing.
BlobShuffle is complementary to these approaches.
Rather than storing operator state remotely, it targets the data exchange between operators and uses object storage as repartitioning layer.
While both lines of work leverage remote storage to improve scalability and elasticity, BlobShuffle focuses specifically on reducing the cost and load of repartitioning traffic.

\paragraph{Messaging over cloud object storage.}
Recently, several---mostly commercial---alternatives to traditional messaging systems such as Apache Kafka have been proposed that rely on cloud object storage as their primary data storage.
Systems such as WarpStream~\cite{WarpStream}, Ursa~\cite{Merli2025}, AutoMQ~\cite{AutoMQ}, and Inkless~\cite{Inkless} use object storage as the main durable log and provide compatibility with the Kafka protocol, allowing existing Kafka clients to connect with little or no modification.
In parallel, there are ongoing activities in the Kafka open-source community~\cite{KIP-1150} to integrate object-storage-backed capabilities directly into Kafka.
At the time of writing, however, it remains unclear if and when such proposals will be adopted in production and what their implications for stream processing workloads will be.

\sloppy %
BlobShuffle shares architectural elements with these object-storage-based messaging systems in that it uses object storage as a durable, cost-efficient substrate for log-like data.
However, these systems typically aim to \emph{replace} existing messaging infrastructures with object-storage-native solutions and thus require changes to the central data backbone used by applications.
BlobShuffle, in contrast, is intentionally less invasive: it integrates on the client side as an add-on to existing stream processing frameworks and uses object storage only as an intermediate shuffle layer.
This design allows applications to retain their existing Kafka (or Kafka-compatible) infrastructure while still benefiting from the cost advantages of object storage for shuffle-heavy workloads.

\section{Conclusions and Future Work}\label{sec:conclusions}

BlobShuffle is a client-side, object-storage-based shuffle mechanism for stream processing systems that groups records into batches, stores these batches in cloud object storage, and forwards only compact notifications via a messaging layer.
By offloading shuffle data from a messaging backbone to object storage, BlobShuffle reduces cross-AZ network costs and operational complexity of high-throughput shuffling.
Our implementation integrates with the popular Kafka Streams framework and requires only minimal code changes to existing applications, while preserving Kafka Streams' processing guarantees.

In an experimental evaluation using a Kubernetes-based deployment on AWS, we show that BlobShuffle can reduce shuffling costs by more than $40\times$ compared to native Kafka Streams repartitioning.
While these savings come at the expense of higher shuffle latency, the resulting p95 latency below 2~seconds remains acceptable for many application scenarios, such as large-scale observability analytics as we run at Dynatrace. 
Within the range of 4--32\,MiB, the batch size provides an effective control parameter to balance the trade-off between cost and latency.
We further evaluate scalability by scaling up to 48~Kafka Streams instances on 24~virtual machines and find that throughput and costs scale near-linearly without encountering a scalability limit in the evaluated range.
The inevitable coherency overhead stemming from the $n$:$n$ shuffling pattern remains manageable for input rates of more than 2\,GiB/s of uncompressed data.

Overall, we conclude that BlobShuffle makes high-throughput, shuffle-intensive workloads economically feasible in cloud environments where cross-AZ traffic dominates operational costs.
We further hope that our findings and methodology help drive ongoing community efforts around leveraging object storage for messaging and shuffling.

Accompanied by extended evaluations, we plan to explore further optimization potential, for example, by improving performance for multi-threaded stream processing and by integrating additional disk-based cache layers to better support fault tolerance and reconfiguration at runtime.
Moreover, we aim to extend BlobShuffle to use cases beyond Kafka Streams, where tighter control over, for instance, the commit process may further simplify the design and enable additional optimizations.

\bibliographystyle{ACM-Reference-Format}
\bibliography{references}

\end{document}